\documentclass[useAMS,usenatbib]{mn2e}
\usepackage{epsf}
\usepackage{amssymb}
\usepackage[usenames]{color}

\def\plotone#1{\centering \leavevmode
\epsfxsize=\columnwidth \epsfbox{#1}}
\def\plottwo#1#2{\centering \leavevmode
\epsfxsize=.99\columnwidth \epsfbox{#1} \hfil
\epsfxsize=.99\columnwidth \epsfbox{#2}}
\def\plotwide#1{\centering \leavevmode
\epsfxsize=1.99\columnwidth \epsfbox{#1}}

\newcommand{\be}{\begin{equation}}
\newcommand{\ee}{\end{equation}}

\title[X-Ray surface brightness fluctuations in the Coma cluster]{X-Ray
  surface brightness  and gas density fluctuations in the Coma cluster}
\author[Churazov et al.]{E.~Churazov,$^{1,2}$ A.~Vikhlinin,$^{3,2}$
  I.~Zhuravleva,$^{1}$ A.~Schekochihin,$^{4}$ \newauthor
  I.~Parrish,$^{5}$ R.~Sunyaev,$^{1,2}$ W.~Forman,$^{3}$  H.~B\"ohringer,$^{6}$ S.~Randall$^{3}$
 \newauthor \\
$^1$ Max-Planck-Institut f\"ur Astrophysik, Karl-Schwarzschild-Strasse 1, 85741
Garching, Germany\\
$^2$ Space Research Institute (IKI), Profsoyuznaya 84/32, Moscow 117997, 
Russia\\
$^3$ Harvard-Smithsonian Center for Astrophysics, 60 Garden St.,
Cambridge, MA 02138, USA \\
$^4$ Rudolf Peierls Center for Theoretical Physics, University of Oxford, Oxford, OX13NP, UK \\
$^5$ Department of Astronomy and Theoretical Astrophysics Center, University of California Berkeley, Berkeley, CA 94720, USA\\
$^6$ MPI f\"{u}r Extraterrestrische Physik, P.O.\ Box 1603, 85740
Garching, Germany\\
}

\begin{document}

\pagerange{\pageref{firstpage}--\pageref{lastpage}}
\pubyear{2009}

\maketitle

\label{firstpage}
\begin{abstract}
X-ray surface brightness fluctuations in the core ($650 \times 650$
kpc) region of the Coma cluster observed with XMM-Newton and
Chandra are analyzed using a 2D power spectrum approach.
The resulting 2D spectra are converted to 3D power spectra
of gas density fluctuations. Our independent analyses of the XMM-Newton and Chandra observations are in excellent
agreement and provide the most sensitive measurements of surface brightness and
density fluctuations for a hot cluster. We find that the
characteristic amplitude of the volume filling density fluctuations
relative to the smooth underlying density distribution varies from 7-10\% on
scales of $\sim$500 kpc down to $\sim$5\% at scales $\sim$ 30 kpc. On smaller
spatial scales, projection effects smear the density fluctuations by a
large factor, precluding strong limits on the fluctuations in
3D.  On the largest scales probed (hundreds of kpc), the dominant contributions to
the observed fluctuations most likely arise from perturbations of the
gravitational potential by the two most massive galaxies in Coma,
NGC4874 and NGC4889, and the
low entropy gas brought to the cluster by an infalling group. Other
plausible sources of X-ray surface brightness fluctuations are discussed, including
turbulence, metal abundance variations, and unresolved sources. Despite a variety of possible origins for density fluctuations, the gas in the Coma cluster core is remarkably homogeneous on scales from $\sim$ 500 to $\sim$30 kpc.
\end{abstract}


%

\sloppypar

\section{Introduction}
The hot intracluster medium (ICM), which fills the cluster
gravitational potential, is the dominant baryonic component of rich
galaxy clusters. In a smooth and static gravitational potential,
the gas in hydrostatic equilibrium should have a smooth density and
temperature distribution aligned with the equipotential surfaces. From
X-ray data both the density and temperature radial profiles are
routinely measured, and the comparison of mass estimates from 
hydrostatic equilibrium and lensing suggests overall good agreement
(to within 10-20\%). However, there are a variety of reasons why the
gas properties may be perturbed on a range of spatial scales:
perturbation of the gravitational potential, turbulent gas motions,
imperfect mixing of the gas with different entropies displaced
  by gas motions, presence of bubbles filled with relativistic
plasma, etc. These perturbations of the ICM properties can be detected
in raw observational data as flux (surface brightness) fluctuations,
variations of temperature, or variations of the projected velocity
field (which can be measured with future high resolution X-ray
calorimeters, like ASTRO-H).

Several previous studies have explored fluctuations in galaxy clusters.
X-ray surface brightness
  fluctuations on arcminute scales, relative to a smooth model, were detected
  for a sample of clusters using the Einstein Observatory data
  \citep{1990ApJ...364..433S}.  Projected
pressure fluctuations in the Coma cluster, using XMM-Newton
observations were studied by
\citet{2004A&A...426..387S}.  Another approach was adopted by \citet{2008ApJ...687..936K},
where the distribution function of the surface brightness fluctuation
was calculated for the A3667 cluster and linked to the 3D density
fluctuations.

Here we present an analysis of the X-ray surface brightness
fluctuations in the core of the Coma cluster using both Chandra and
XMM-Newton observations, calculate the power spectrum
of these fluctuations and estimate the amplitudes of the 3D density
fluctuations on scales ranging from 30 kpc to 500 kpc. The projected
pressure fluctuations have been analyzed by 
\citet{2004A&A...426..387S} over a limited range of spatial
scales. Using X-ray surface brightness instead of pressure allows one
to probe smaller spatial scales and greatly simplifies 
accounting for statistical noise.
  
In our work the choice of the Coma cluster was motivated by three factors:
proximity (thus allowing us to probe small physical scales), presence of
a large flat surface brightness X-ray core (simplifying the conversion of 2D power
spectra to 3D) and brightness of the cluster. The extension of the analysis to
even brighter and closer clusters, in particular, Perseus and M87, is
straightforward and will be presented elsewhere.

Throughout the paper the redshift of Coma was assumed to be $z=0.023$,
corresponding to an angular diameter distance of 93 Mpc (for
$h=0.72$); $1''$ corresponds to 0.45 kpc.

This paper is organized as follows: in \S\ref{sec:data} we
describe the data-sets used in the analysis. In \S\ref{sec:3d2d} the
relation between 2D and 3D power spectra is explained. The
measurements of the density fluctuation power spectrum are presented
in \S\ref{sec:mes2d}. The implications of our results are discussed in
\S\ref{sec:discussion} and \S\ref{sec:conclusions} contains our
conclusions.

\section{Data and initial processing}
\label{sec:data}
For our analysis we used publicly available XMM-Newton and Chandra
data.

For Chandra, the following OBSIDs were used: 555, 556, 1086, 1112,
1113, 1114, 9714, 10672. The data were prepared following the
procedure described in \cite{2005ApJ...628..655V}. This includes
filtering of high background periods, application of the latest
calibration corrections to the detected X-ray photons, and
determination of the background intensity in each observation. The
total exposure of the resulting data set is $\sim 1.15\times 10^5$ s.

For XMM-Newton, we used a large set of pointings, covering a field of more
than $1\times 1$ degrees (although only the most central part of this
field was used in the subsequent analysis). The data were prepared by
removing background flares using the light curve of the detected
events above 10 keV and re-normalizing the ``blank fields'' background
to match the observed count rate in the 11-12 keV band.  In the
subsequent analysis we used the data from the EPIC/MOS (European Photon
Imaging Camera/Metal Oxide Semi-conductor) detector. The total
exposure of all used observations is 880 ksec, covering a much larger
area than the Chandra observations (not only the cluster core but also the
outskirts of the cluster). 

The images in the 0.5-2.5 and 0.5-4 keV bands were generated for
XMM-Newton and Chandra respectively, along with the exposure and
background maps, which were used for subsequent analysis.

\begin{figure}
\plotone{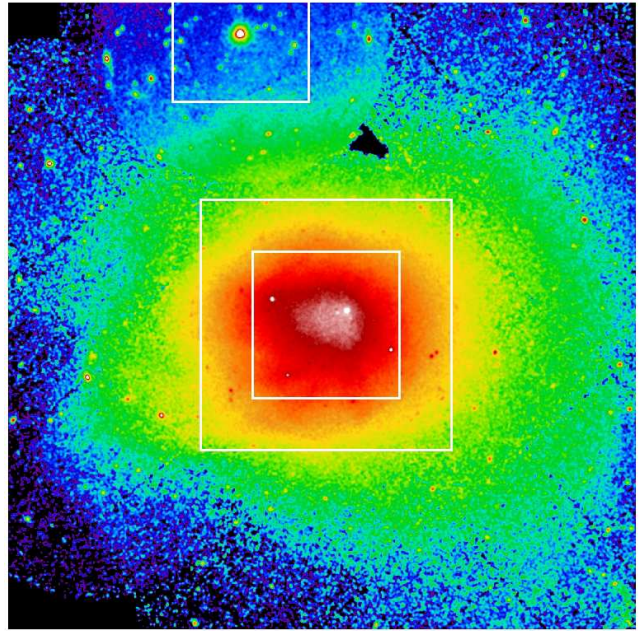}
\caption{XMM mosaic image of Coma ($1^\circ \times 1^\circ$, or
  1.6$~\times~$1.6 Mpc). Boxes show the regions used for the power
  spectrum calculation: central $14' \times 14'$ box - Chandra field;
  larger $24' \times 24'$ box - XMM-Newton field. The smallest $13' \times
  13'$ box near the top of the mosaic, which contains the X-ray bright
  Seyfert galaxy X-Comae, was used to determine the shape of the power
  spectrum produced by point sources in XMM-Newton images.
\label{fig:rawimage}
}
\end{figure}

The background-subtracted, exposure-corrected XMM-Newton image
($1^\circ \times 1^\circ$, or 1.6$~\times~$1.6 Mpc) is shown in
Fig.\ref{fig:rawimage}. For the subsequent analysis of XMM-Newton
  data only the central $24' \times 24'$ part of the image was used
  (roughly corresponding to the flat surface brightness cluster core). For Chandra we also excluded
  edges of the image with low exposure, keeping only the central
  $14' \times 14'$ region.

\begin{figure}
\plotone{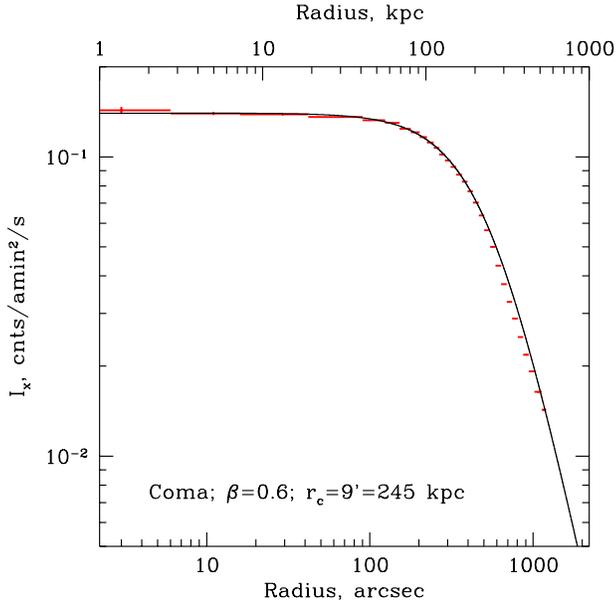}
\caption{Radial X-ray surface brightness profile (XMM-Newton) of the
  Coma cluster. The solid line shows the surface brightness for
  a $\beta$-model with core radius $r_c=9'$ (or $\sim$245 kpc)
  and $\beta=0.6$.
\label{fig:sbraw}
}
\end{figure}

An azimuthly averaged X-ray surface brightness profile $I(R)$ is shown
  in Fig.\ref{fig:sbraw}. The observed distribution can be reasonably
  well described by a simple $\beta$-model \citep{1978A&A....70..677C} with core radius $r_c=9'$ (or
  $\sim$245 kpc) and $\beta=0.6$ 
\begin{eqnarray}
I(R)=\frac{I_0}{\left [ 1+ \left ( \frac{R}{r_c}\right )^2\right ]^{3\beta-0.5} },
\label{eq:sbbeta}
\end{eqnarray}
where $R$ is the projected distance from the cluster center and $I_0$ is the surface brightness at the center.
In \S\ref{sec:div} we
  use this model (and a more complicated one) to remove the global
  cluster emission. As discussed in \S\ref{sec:simple} the results of
  our analysis are only weakly sensitive to the particular choice of
  the $\beta$-model. Partly this is because our analysis is applied to
  the core region of the Coma cluster. 

\section{3D to 2D projection}
\label{sec:3d2d}

We assume that the density distribution in the cluster is
described by a simple decomposition into undisturbed and fluctuating components
\begin{eqnarray}
n(x,y,z)=n_{0}(x,y,z)\left [1+\delta(x,y,z) \right ],
\end{eqnarray}
where $\displaystyle n_{0}(x,y,z)$ is the undisturbed density
distribution (e.g. standard spherically symmetric
  $\beta$-model\footnote{One can imagine more complicated models,
    e.g. ellipsoidal $\beta$-model or a model with more sophisticated radial
  density profiles. In \S\ref{sec:discussion} we argue that the freedom in choosing the
  undisturbed model mostly affects the power spectra at large
  scale.} shown in Fig.\ref{fig:sbraw} and the dimensionless quantity
$\displaystyle \delta(x,y,z) = \frac{\delta n}{n_0}\ll 1$ describes the fluctuating part.
For ICM temperatures above $\sim$3 keV, the emissivity of 
  plasma in the soft band, which dominates the flux detected by
  XMM-Newton and Chandra, mainly depends on the square of the electron
  density and is only weakly sensitive to the temperature\footnote{The
    impact of possible abundance variations is briefly discussed in \S\ref{sec:metalls}}. 
Accordingly the volume emissivity $\varepsilon$ can be written as 
\begin{eqnarray}
\varepsilon(x,y,z)= C n^2(x,y,z)\approx
Cn^2_{0}(x,y,z)\left[1+2\delta(x,y,z)\right],
\label{eq:ne0}
\end{eqnarray}
neglecting terms of order $\delta^2$  and $C$ is the coefficient
relating the square of the density and the emissivity. Since we are
interested in fluctuations, relative to the average surface brightness
profile, for the subsequent analysis we set $C=1$. We further assume that
$\delta(x,y,z)$ is a homogeneous and isotropic random field,
characterized by the power spectrum $P_{3D}$:
\begin{eqnarray}
P_{3D}(k_x,k_y,k_z)=\left |\int
\delta(x,y,z)e^{-i2\pi(xk_x+yk_y+zk_z)}dxdydz \right |^2.
\label{eq:3d}
\end{eqnarray}
By isotropy assumption, $\displaystyle
P_{3D}(k_x,k_y,k_z)=P_{3D}(|k|)$, where $k=\sqrt{k_x^2+k_y^2+k_z^2}$. 

The observed image $I(x,y)$ is the integral of the emissivity along the
line of sight:
\begin{eqnarray}
I(x,y)=\int \varepsilon(x,y,z) dz =\\ \nonumber
\int n^2_{0}(x,y,z) dz + 2\int n^2_{0}(x,y,z)\delta(x,y,z) dz =
\\ \nonumber
I_0(x,y)+2\int n^2_{0}(x,y,z)\delta(x,y,z) dz,
\end{eqnarray}
where $I_0(x,y)$ is the undisturbed surface brightness.
Dividing the observed image by $I_0(x,y)$ yields the normalized image $J(x,y)$
\begin{eqnarray}
J(x,y)=1+2\frac{\int n^2_{0}(x,y,z)\delta(x,y,z) dz}{\int
  n^2_{0}(x,y,z)dz}.
\label{eq:j}
\end{eqnarray} 
The 2D power spectrum of $J(x,y)$ 
\begin{eqnarray}
P_{2D}(k_x,k_y)=\left | \int J(x,y)e^{-i2\pi(xk_x+yk_y)}dxdy \right |^2
\end{eqnarray} 
can be 
  obtained directly from
observational data. By isotropy, $\displaystyle
P_{2D}=P_{2D}(k_{xy})$, where $\displaystyle
k_{xy}=\sqrt{k_x^2+k_y^2}$. The expression (\ref{eq:j}) contains the
following function of $x$, $y$ and $z$ under the integral
\begin{eqnarray}
\eta(x,y,z)=\frac{n^2_{0}(x,y,z)}{\int
  n^2_{0}(x,y,z')dz'}=\frac{n^2_{0}(x,y,z)}{I_0(x,y)}.
\end{eqnarray} 
For the range of projected distances we consider, the weak dependence
of $\eta(x,y,z)$ on $x$ and $y$ can be neglected (the changes in
$\eta(x,y,z)$ are at most
factor of 2 across the whole image $24' \times 24'$ for the adopted $\beta$-model). Thus $\eta(x,y,z)\approx \eta(z)$ and
the $P_{2D}$ spectrum can be linked to the $P_{3D}$ of
$\delta(x,y,z)$ as follows (\citealt{1999coph.book.....P}, see also \citealt{iz11})
\begin{eqnarray}
P_{2D}(k_{xy})=4\int{P_{3D}\left (\sqrt{k_{xy}^2+k_z^2} \right )|W(k_z)|^2dk_z},
\label{eq:32}
\end{eqnarray}
where  $|W(k_z)|^2$ is
the 1D power spectrum of the normalized emissivity distribution
along the line of sight. 
\begin{eqnarray}
W(k_z)=\int \eta(z) e^{-i2\pi z k_z} dz.
\label{eq:w}
\end{eqnarray} 

\begin{figure}
\plotone{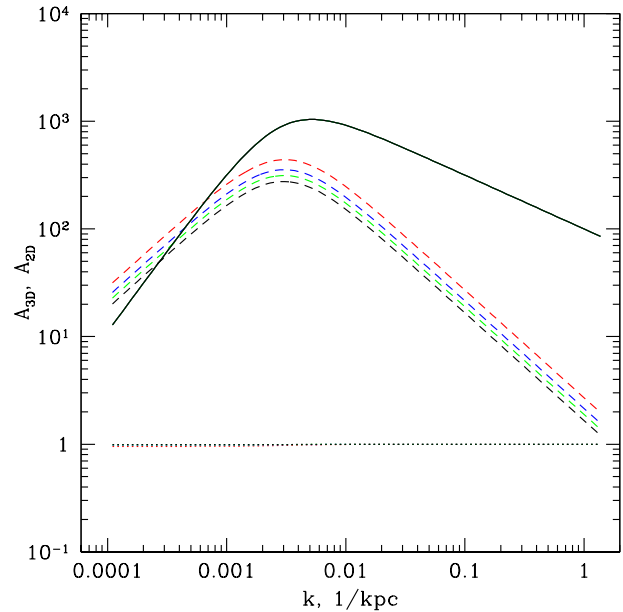}
\caption{3D (solid) and 2D (dashed) amplitudes as a function of $k$
  for a model power spectrum. The 3D power spectrum has $k_m=3~10^{-3}~{\rm kpc^{-1}}$
  and $\alpha=4$.  The 2D amplitude is calculated for 
 different projected distances $R=$0, 300, 435, 600 kpc (red, blue, green, black
 respectively). The dotted line shows the ratio
  of $A_{2D}$ to the prediction of the simplified expression
  (\ref{eq:klarge}).
\label{fig:p32los}
}
\end{figure}

Note that the properties of the emission measure distribution along
the line of sight determine the characteristic wavenumber $k_{z,cutoff}$
above which $|W(k_z)|^2$ falls off towards larger $k$. For $k\gg k_{z,cutoff}$ the
expression (\ref{eq:32}) simplifies to
\begin{eqnarray}
P_{2D}(k)\approx 4P_{3D}(k)\int{|W(k_z)|^2dk_z},
\label{eq:klarge}
\end{eqnarray}
i.e., the 2D power spectrum of the surface brightness fluctuations is
essentially equal to the 3D power spectrum of the density fluctuations
apart from the normalization constant $4\int{|W(k_z)|^2dk_z}$, which
is easily measured for a cluster. In principle $\int{|W(k_z)|^2dk_z}$ depends on $x$ and $y$ and so is different for different lines of sight, but
as we show in Fig.\ref{fig:p32los}, for the central region of Coma this
difference can be neglected. 

For practical reasons it is easier to use the characteristic amplitude of
the fluctuations, rather than the power spectrum. The amplitudes corresponding to the 3D
and 2D spectra are defined as 
\begin{eqnarray}
A_{3D}(k)= \sqrt{P_{3D}(k)4\pi k^3},
\label{eq:a3}
\end{eqnarray}
\begin{eqnarray}
A_{2D}(k)= \sqrt{P_{2D}(k)2\pi k^2}
\label{eq:a2}
\end{eqnarray}

To illustrate the above relations for the case of Coma, we set the density
distribution eq.(\ref{eq:ne0}) to the $\beta$-model with $\beta=0.6$
and core radius $r_c=245$ kpc corresponding to the azimuthally
  averaged Coma surface brightness profile, shown in Fig.\ref{fig:sbraw}: 
\begin{eqnarray}
n_0(r)\propto \left (1 + \frac{r^2}{r_c^2} \right)^{-\frac{3}{2}\beta}
\end{eqnarray}
and calculated the power spectra of the
line-of-sight emissivity distribution at several projected radii. We
further parametrize the 3D power spectrum of density fluctuations with a simple cored power
law 

\begin{eqnarray}
P_{3D}(k)=\frac{B}{\left ( 1+\frac{k^2}{k_m^2}\right )^{\alpha/2}},
\label{eq:psmodel}
\end{eqnarray}
where $\alpha$ is the slope of the power spectrum at large $k$,
where $k_m$ is the characteristic wavenumber below which the 3D
spectrum is flat and $B$ is the normalization constant. This
  choice of the power spectrum parametrization is rather arbitrary. It
  is used only to compare the exact calculation of  $P_{2D}(k)$ using
  eq.(\ref{eq:32}) with the approximate formula
  eq.(\ref{eq:klarge}).

\begin{figure}
\plotone{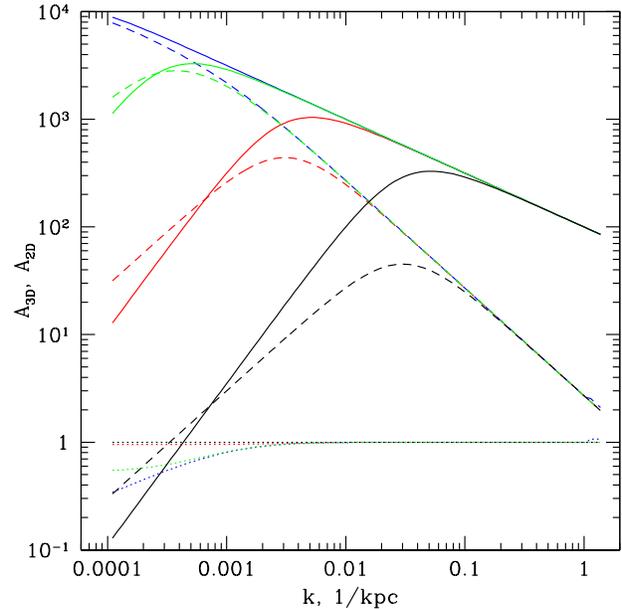}
\caption{3D and 2D amplitudes as a function of $k$ for a model power
  spectrum and for a $\beta$-model cluster with $\beta=0.6$ and
  $r_c=245$ kpc. A cored power law model is used for the 3D power
  spectrum, according to eq.(\ref{eq:psmodel}). Different colors
  correspond to different break wavenumber (in ${\rm kpc^{-1}}$)  in the 3D power spectrum:
  $k_m=3~10^{-5}$ (blue), $3~10^{-4}$ (green), $3~10^{-3}$ (red),
  $3~10^{-2}$ (black). Projection (see
  eq.(\ref{eq:32})) is done along the line-of-sight going through
the Coma center. Solid curves correspond to $A_{3D}$, dashed -- to
$A_{2D}$, dotted -- to the ratio of $A_{2D}$ to the prediction of the
simplified expression (\ref{eq:klarge}). All dotted curves are
  very close to unity at $k\ge 10^{-3}~{\rm kpc^{-1}}$, indicating
  that  expression (\ref{eq:klarge}) can be used to relate 3D and 2D
  amplitudes for $k\ge 10^{-3}~{\rm kpc^{-1}}$.
 In this plot, the factor 4
appearing in eq. (\ref{eq:32}) is omitted.
\label{fig:p32km}
}
\end{figure}

\begin{figure}
\plotone{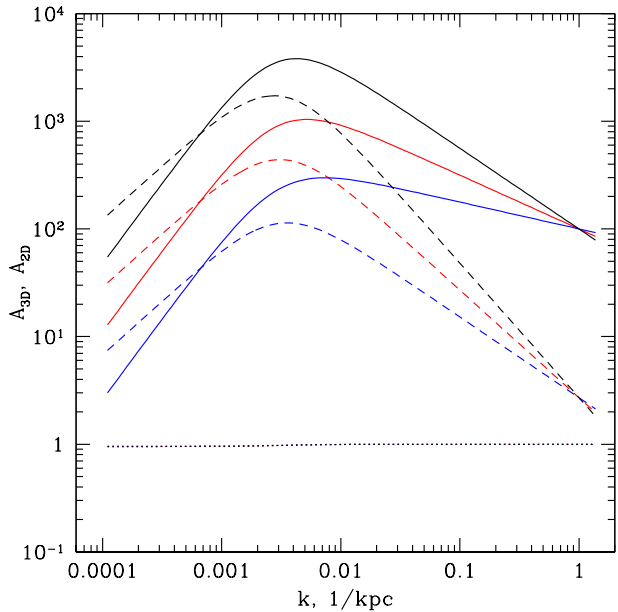}
\caption{3D (solid) and 2D (dashed) amplitudes as a function of $k$
  for a model power spectrum. Different slopes of the 3D power spectrum
  $\alpha=$3.5, 4, 4.5 are shown with blue, red and black colors
  respectively. $k_m=3~10^{-3}~{\rm kpc^{-1}}$ is the same for all
  plots. The dotted line
  shows the ratio of $A_{2D}$ to the prediction of the simplified expression
  (\ref{eq:klarge}). 
\label{fig:p32slope}
}
\end{figure}

Using eq. (\ref{eq:psmodel}) and the power spectrum $|W(k_z)|^2$, the
power spectrum of the surface brightness fluctuations can be easily
calculated via eq. (\ref{eq:32}). The resulting characteristic
amplitudes $A_{3D}$ and $A_{2D}$ for different values of $k_m$,
$\alpha$ and different lines of sight (projected distances) are shown
in Figs. \ref{fig:p32km}, \ref{fig:p32slope} and \ref{fig:p32los}
respectively. 

In each of these images the dotted lines show the ratio of the 2D
  and 3D amplitudes, calculated using the exact relation (\ref{eq:32})
  divided by the ratio evaluated using the simplified relation
  (\ref{eq:klarge}), which corresponds to the case of small scale
  perturbations. Clearly, independently of the shape of the power
  spectrum (Fig.\ref{fig:p32km} and Fig.\ref{fig:p32slope}) or
  projected distance (Fig.\ref{fig:p32los}) the simplified relation
  (\ref{eq:klarge}) provides an excellent approximation of the
  relation between 2D and 3D spectra (unless the power law spectrum
  extends to very low $k$, corresponding to scales of a few
  Mpc). Furthermore for a given 3D spectrum, the expected 2D spectra,
  measured at different projected distances from the Coma cluster,
  differ only in amplitude, and this difference is less than a factor
  of 2 (provided that the projected distance is less than $\sim600$
  kpc).

\section{Measuring the 3D power spectrum of density fluctuations}
\label{sec:mes2d}

Our goal is to obtain 3D power spectra of the density
fluctuations. Our general strategy is as follows. We
start with the analysis of the 2D power spectra of raw images and
evaluate the contribution of the Poisson noise and point sources. Then
we pick a smooth model as a representation of the undisturbed cluster
density distribution, predict the corresponding surface
  brightness distribution  
and calculate 2D fluctuations relative to this model. The resulting
2D power spectra are corrected for the telescope Point Spread Function (PSF) effects. We then convert
2D power spectra to 3D spectra using eq.(\ref{eq:klarge})
and recast the results in terms of 3D amplitudes of density
fluctuations relative to the smooth 
  underlying 3D model.

\subsection{2D power spectrum evaluation}

\begin{figure}
\plotone{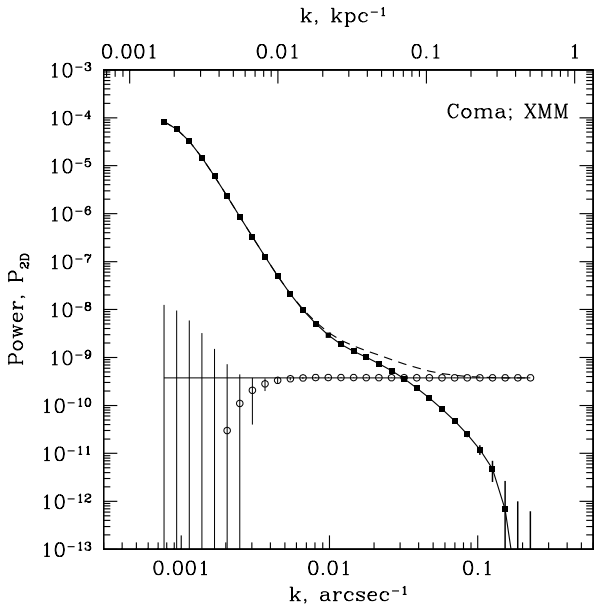}
\caption{Raw power spectrum $P_{2D}(k)$ (dashed line), eq.(\ref{eq:rms2k}),  of the XMM-Newton field with
  sources and Poisson noise. The Poisson noise level, estimated
  using eq.(\ref{eq:poi}), is shown with the thin horizontal line. The Poisson
  noise evaluated by adding extra Poisson noise to the images (see
  \S\ref{sec:poi}) is shown with circles. The Poisson-subtracted power spectrum is shown with the solid line and black
  squares. The errors shown are due to Poisson noise only and do not
  account for the stochastic nature of the signal itself.
\label{fig:psraw}
}
\end{figure}

The method used to evaluate low resolution power spectra of the
surface brightness fluctuations is described in \citealt{Are11} (see
also \citealt{2008A&A...485..917O}). Briefly one convolves the image
with a set of Mexican-Hat filters ( with different spatial scales) $\displaystyle F(x)= \left [1 -
  \frac{x^2}{\sigma^2} \right ] e^{-\frac{x^2}{2\sigma^2}}$, where
$\sigma$ is the characteristic spatial scale\footnote{Mexican-Hat
  filter has a positive core and negative wings, and preferentially
  selects perturbations with a given scale.},
 and relates the variance of the resulting image to the power at
a given scale. The presence of a non-uniform exposure map or gaps in
the data (e.g. excised regions around bright point sources) is
accounted for by representing the Mexican Hat filter as a difference
between two Gaussians with only slightly different width.  For each
spatial scale $\sigma$, the following simple steps are performed (see
\citealt{Are11} for a detailed description):
\begin{description}
\item[(i)] the observed image $I$ and the exposure map $E$ are smoothed with
  two Gaussians $G$ having different smoothing lengths
  $\sigma_1=\sigma/\sqrt{1+\epsilon}$ and
  $\sigma_2=\sigma\sqrt{1+\epsilon}$, where $\epsilon \ll 1$ is a
  dimensionless number;
\item[(ii)] smoothed images are divided by smoothed exposure maps and
  regions with low (unsmoothed) exposure are masked out;
\item[(iii)] the difference of the two images is calculated, this
  difference image is dominated by fluctuations at scales $\sim \sigma$;
\item[(iv)] the variance of the resulting image is calculated and
  re-cast into an estimate of the power, using a normalization
  coefficient that depends on $\epsilon$. 
\end{description}
Thus the variance is calculated for an image
\begin{equation}
I_{\sigma}=E\times \left (
\frac{G_{\sigma_1}\circ I_{raw}}{G_{\sigma_1}\circ E}-\frac{G_{\sigma_2}
  \circ I_{raw}}{G_{\sigma_2} \circ E}
\right ),
\label{eq:isigma}
\end{equation}
where $E=E(x,y)$ is the exposure map, $I_{raw}=I(x,y)$ is the raw
image in counts and $\displaystyle G_{\sigma}\circ I$ denotes
a convolution of an image $I$ with a Gaussian. The final expression for the
power at a given spatial scale (or given wavenumber) is then 
\begin{equation}
P_{2D}(k)=\frac{1}{\epsilon^2\pi k^2}\frac{\Sigma I_{\sigma}^2}{\Sigma E^2},
\label{eq:rms2k}
\end{equation}
where summation is over the image pixels \citep[see appendix 1 in][]{Are11}.
The resulting 2D power spectrum for the XMM-Newton image of Coma is plotted in
Fig.\ref{fig:psraw} (dashed line). 
Effectively the power spectrum shown in Fig.\ref{fig:psraw} 
is a convolution of the true underlying power spectrum with a broad ($\Delta k \sim
k$)
filter in $k$ space. Therefore individual measurements of the power are not
independent, unlike components of the conventional Fourier transform in a
rectangular region. However the procedure adopted here has a clear
advantage compared to the conventional Fourier transform, since most of the problems
caused by non-periodicity of the data-set and the presence of the gaps
in data are automatically corrected for \citep{Are11,2008A&A...485..917O}. 

For XMM-Newton and Chandra observations, we decided to use images
without instrumental and cosmic background subtraction. Inclusion of the background does not
significantly affect the results, but in the case of XMM-Newton 
  it would add some
high frequency noise due to weak compact sources, which were not eliminated
completely in the blank-field data-sets.

\subsection{Poisson noise}
\label{sec:poi}

The Poisson noise contribution can be easily estimated for the raw
images. We expect a white noise *(flat) power spectrum* for this
component with the power:
\begin{equation}
P_{Poisson}=\frac{\Sigma I_{raw}}{\Sigma E^2},
\label{eq:poi}
\end{equation}
where summation is over the image pixels.
The result of this calculation is shown in Fig.\ref{fig:psraw} with the
solid horizontal line.

Very small deviations from this simple estimate are possible due to
non-trivial behavior of the filter at the edges of the image or near
the data gaps (partly due to the coarse representation of the Mexican Hat
filter on a grid for the highest frequency bins). To take care of
these possible deviations we made 100 realizations of fake raw images by
adding extra Poisson noise to each pixel with the mean value set by
the actual number of counts in the pixel. The whole procedure of
evaluating the power spectrum is then repeated and the mean of the
resulting power spectra is calculated. Effectively this mean power
spectrum has an extra Poisson noise component (compared to the
original image) with all properties similar to the Poisson noise in the original
image. Therefore the difference between the mean power spectrum and the original
power spectrum provides an estimate of the Poisson noise contribution,
which includes all possible effects of the filter implementation. 

 As expected, the derived noise spectrum has an essentially flat spectrum. 
The same procedure allows us to evaluate
uncertainties caused by the presence of Poisson noise in the power
spectrum. The estimated level of
the Poisson noise and corresponding uncertainties (in evaluating
the Poisson noise) are shown in  Fig.\ref{fig:psraw} as black 
circles with error bars. 

The power spectrum with the Poisson noise contribution subtracted is shown
with the solid curve. The errorbars shown are due to the Poisson noise
only.

\begin{figure*}
\plottwo{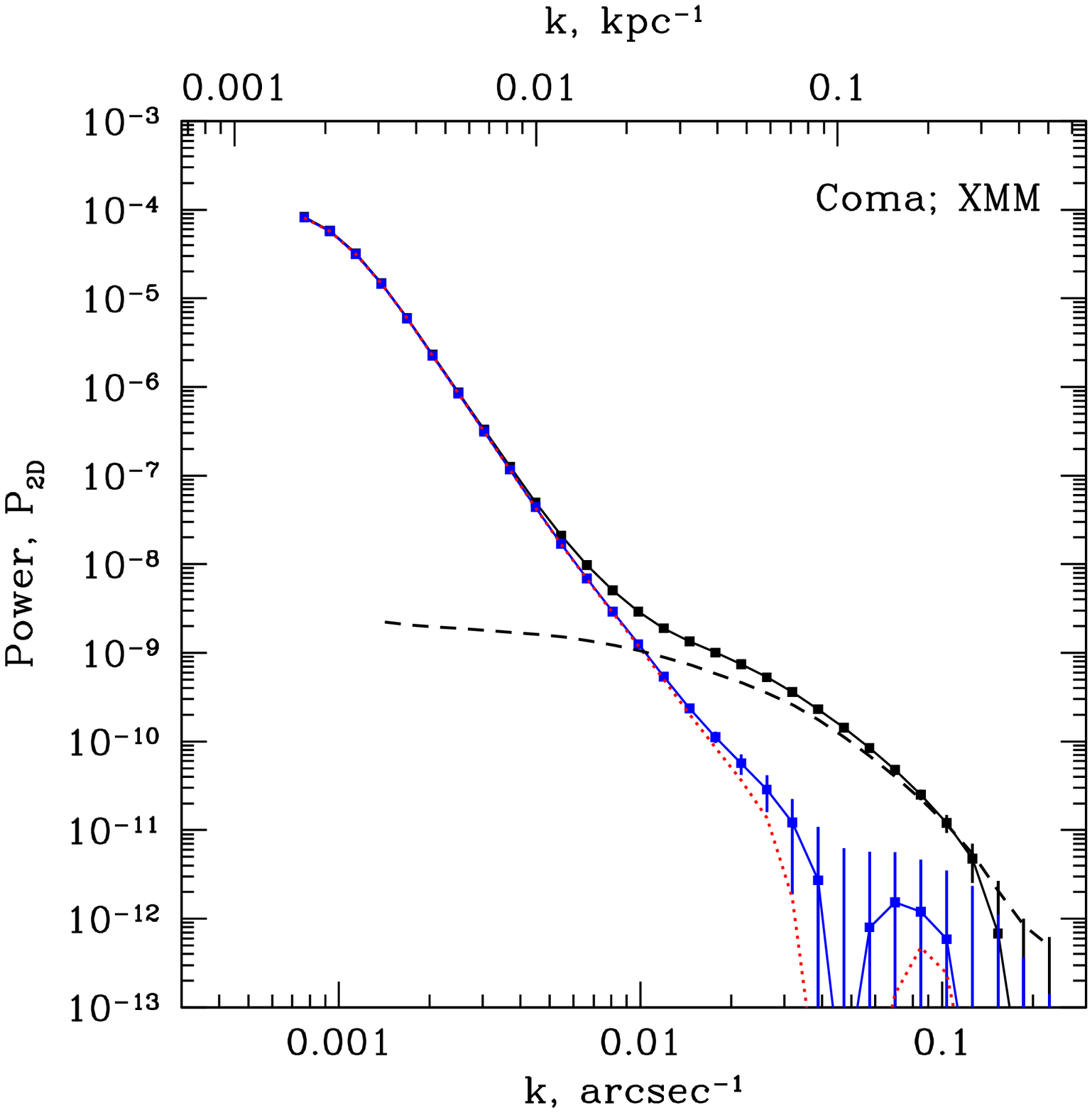}{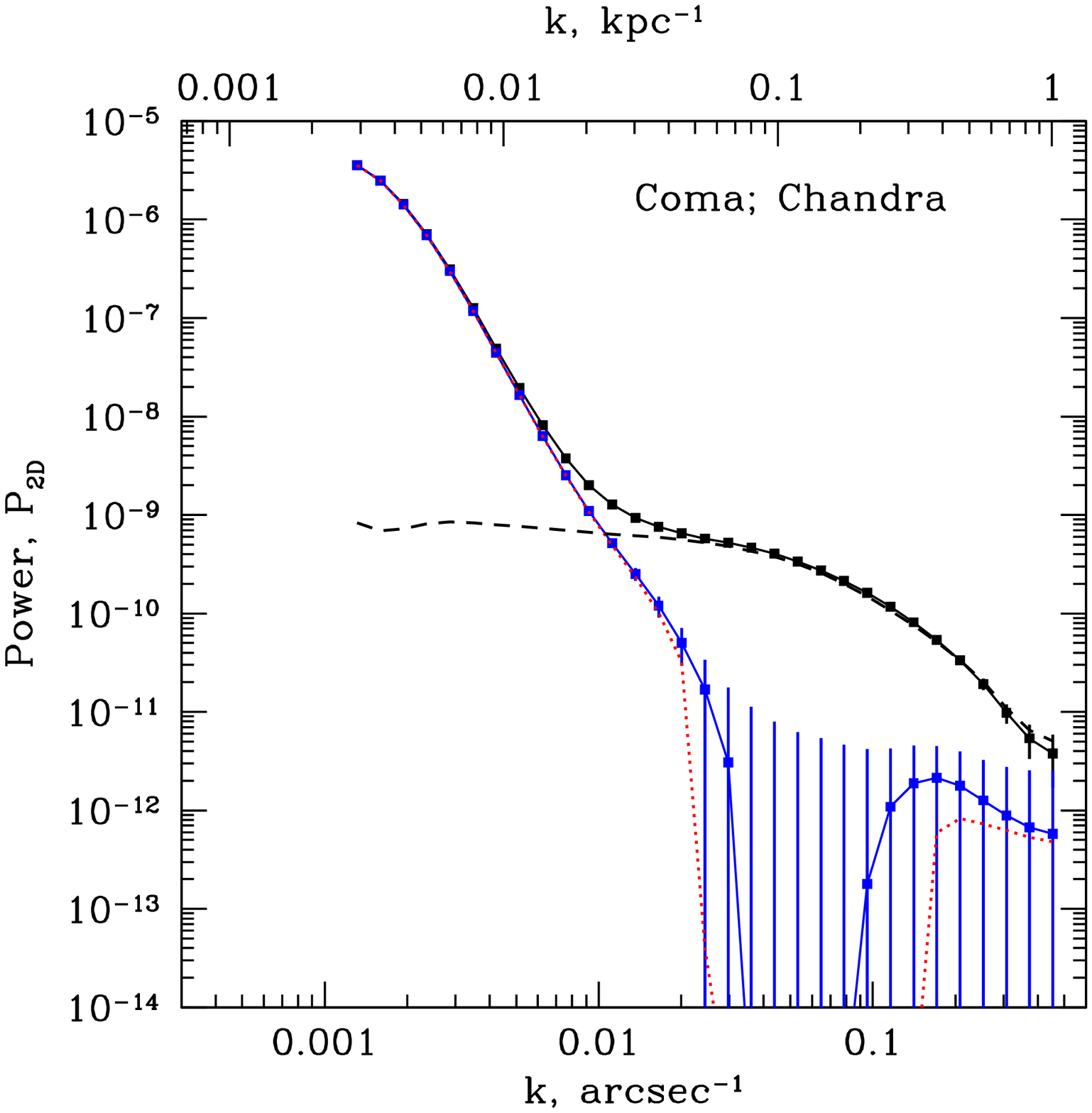}
\caption{Power spectra of raw images with point sources (black solid) and with $15''$ circles around
 point sources excluded (blue solid) for XMM-Newton field (left) and
 Chandra field (right). Black dashed lines show the expected shape of
 a power spectrum associated with point sources, taken from the
 X-Comae and Lockman fields for XMM-Newton and Chandra
 respectively (see \S\ref{sec:src}). Dotted red lines show the
 effect of subtracting an extra component, associated with weak point
 sources, which are too faint to be detected individually in the
 images. The normalization of this subtracted component is estimated to be
 $\approx$3\% of the power spectrum due to bright detected sources (see \S\ref{sec:src}).
\label{fig:prsrc}
}
\end{figure*}

\subsection{Point sources}
\label{sec:src}
The Poisson-noise-subtracted power spectrum (solid line in
Fig.\ref{fig:psraw}) changes slope at $k\sim 0.01~{\rm arcsec^{-1}}$. 
At larger $k$, the spectrum flattens. This  flattening is at least partly
caused by the presence of compact sources in the image. Indeed,
ignoring the blurring of an image by the PSF of the telescope, the presence of compact sources
(i.e. $f_{src}\delta(x-x_{src},y-y_{src})$, where $\delta$ is the
Dirac delta-function at the position of a source with flux $f_{src}$),
randomly scattered across the image, produces a flat power spectrum
over the entire range of wavenumbers. Accounting for the telescope
  PSF, having Gaussian width $\sigma$, the expected contribution of point sources has the shape of the PSF
power spectrum, which is flat at $k\ll 1/\sigma$ and falls off at
larger wavenumbers.

As a first step towards removing the contribution of compact sources
we excised a $15''$ circle (radius) from the images
around each bright, individually detected source. The exposure map has been modified
accordingly. The power spectra calculated from the images without
bright sources are shown in Fig.\ref{fig:prsrc} with the blue
curves. Clearly, much of the high-frequency part of the power spectrum
goes away once bright sources are excluded. For comparison, the raw
spectrum is shown with black symbols.

To confirm still further that much of the detected higher-frequency
power is attributable to the point sources, we computed the power
spectrum for images containing a large number of bright compact sources.

\begin{figure}
\plotone{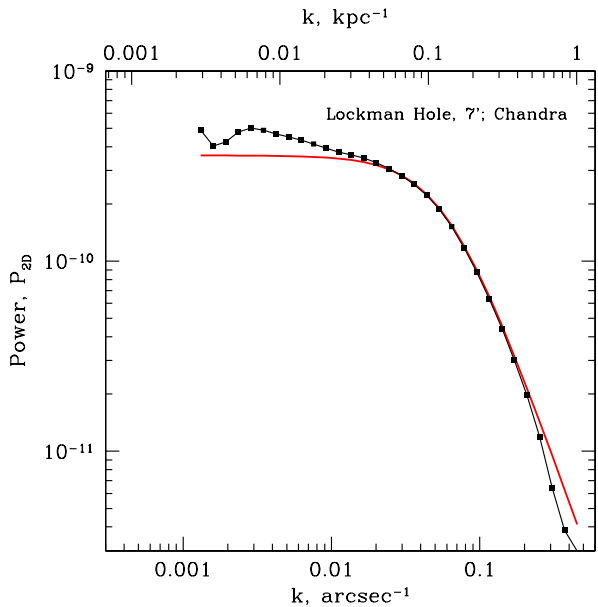}
\caption{Power spectrum (black) of the Chandra Lockman Hole image.
  The red curve shows the crude approximation of the power
  spectrum of the Lockman Hole field by a simple model (see
  \S\ref{sec:psf} and eq.\ref{eq:psf})
\label{fig:lockps}
}
\end{figure}

For Chandra, we used one of the long observations of the Lockman Hole
with ACIS-I and used a $14'\times 14'$ patch of the image.
 The power spectrum obtained from this field after
correction for Poisson noise is shown in Fig.\ref{fig:lockps}. We also plot
the same power spectrum in Fig.\ref{fig:prsrc} with a dashed line, changing
only the normalization of the spectrum. Clearly the shape of the
shoulder in the power spectrum calculated for a raw image of Coma
(including compact sources) is consistent with the power spectrum observed
in the Lockman Hole data.

\begin{figure}
\plotone{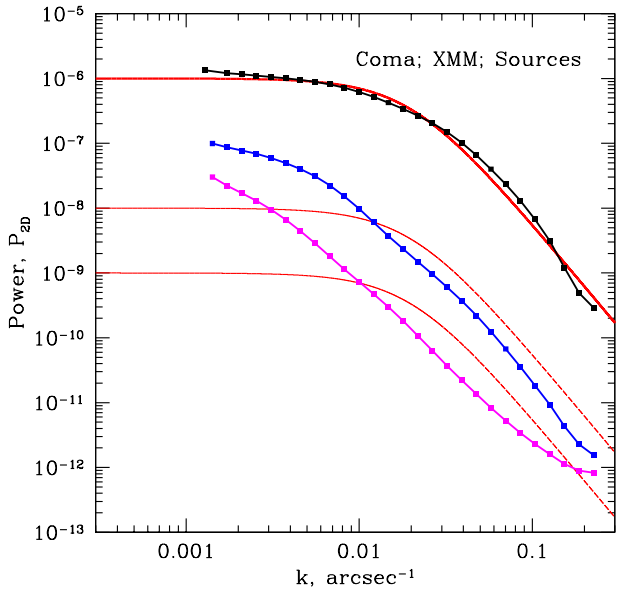}
\caption{Power spectra for a patch near X Comae. Black curve - the
  power spectrum from the image with sources (see Fig.\ref{fig:comae},
  left panel). Blue - power spectrum when $15''$ circles (radius) around bright sources
  are excluded (see Fig.\ref{fig:comae},
  middle panel). Magenta - power spectrum when $30''$ circles excluded (see Fig.\ref{fig:comae},
  right panel). Thick red curve shows a simple approximation of
  the PSF power spectrum (see \S\ref{sec:psf} and
  eq.\ref{eq:psf}). Lower red curves show the same spectrum scaled down
  by a factor of 100 and 1000. From this plot one can conclude that
  excising $15''$ circle around a bright source reduces the contribution of
  the source to the power spectrum by a factor of $\sim$300 or more (see \S\ref{sec:src}).
\label{fig:xsrc}
}
\end{figure}

For XMM-Newton, we used a patch of the image to the NE from the Coma
center (Fig.\ref{fig:rawimage}) . This patch contains the X-ray
bright Seyfert galaxy X-Comae and a number of weaker sources. The
power spectrum calculated for this image is shown in
Fig.\ref{fig:xsrc}.  Similarly to Chandra, this power spectrum has a
shape consistent with the shape of the shoulder obtained from the raw
image of Coma (including compact sources) as shown in
Fig.\ref{fig:prsrc}. The properties of the PSF vary across the
  telescope field of view. We used a field close to the Galactic
  Plane (OBSID: 0605580901), containing many compact sources across the field, to verify
  that the shape of the PSF power spectrum obtained from the X-Comae
  field is not too strongly dominated by the X-Comae itself, which was
  observed close to the center of the XMM-Newton field of view. The
  resulting power spectrum of the Galactic Plane image is consistent
  with the X-Comae patch over the interesting range of wavenumbers.

\begin{figure*}
\plotwide{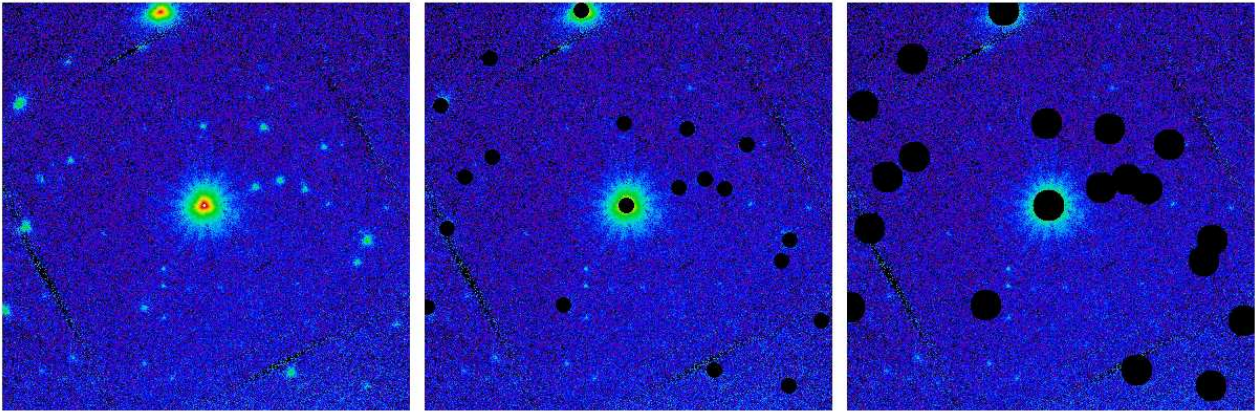}
\caption{Patch of the XMM-Newton image ($13'\times 13'$) near the
  bright Seyfert galaxy X Comae. In the middle
  and right panels the circles with radius $15''$ and $30''$ are excised
  around bright sources.
\label{fig:comae}
}
\end{figure*}

An additional question arises in connection with XMM-Newton data: is
a $15''$ circle around a source large enough to sufficiently suppress the
contribution to the Power Spectrum of the PSF wings? To address this question,
we repeated the calculation of the power spectrum for the patch
near X-Comae, excluding $15''$ and $30''$ circles around each source
(see Fig.\ref{fig:comae}). The corresponding power spectra are shown in
Fig.\ref{fig:xsrc}. From the comparison of these spectra with the full
power spectrum of the X-Comae patch, it is clear that removing $15''$
circles suppresses the high frequency noise by a factor of order $300$, while
$30''$ circles suppress the signal\footnote{Note that there is significant contribution of Coma
  cluster diffuse emission to the field. For this reason the signal
  does not go to zero even for the larger size of the excised
  regions.} by a factor of a few
1000. For our purposes, the suppression factor of more than a 100
is enough and we fixed the size of the excised region
around each point source to $15''$ (radius).


\begin{figure}
\plotone{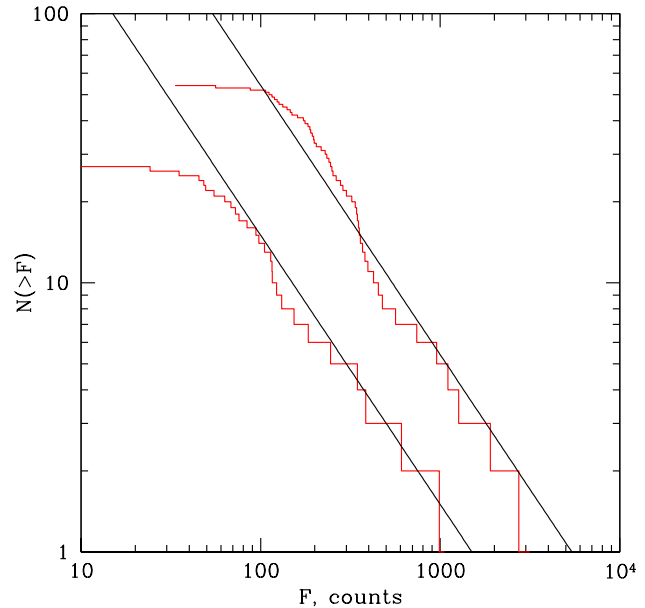}
\caption{LogN-LogS for detected sources in XMM and Chandra fields
  (upper and lower red histograms, respectively). The
  flux (in counts) for each source was calculated in $15''$ circle,
  using $30''$ for the background. Black solid lines show the $F^{-1}$
  law. Assuming that the same law continues to lower fluxes, one can
  estimate the contribution of faint (undetected) sources to the power
  spectrum (see \S\ref{sec:src}).
\label{fig:unresolved}
}
\end{figure}

Apart from the bright detected sources there must be faint compact
objects that we failed to detect. Their collective contribution can
be estimated assuming that we know the LogN-LogS and limiting fluxes. 
To this end we calculated fluxes of our detected sources using a $15''$
circle as a source region and a twice larger circle as a background
region for each source (for Chandra the size of the region is probably
larger than necessary, but it is not very important for the present
study). The LogN-LogS curves for both instruments are shown in
Fig.\ref{fig:unresolved} (note, that the XMM field is larger than the
Chandra field). The curves can be reasonably well approximated by the
law $N\propto F^{-1}$. Assuming that the same slope continues at lower
fluxes we can estimate the contribution of faint point sources to the
power spectrum:
\begin{equation}
P_{faint}\propto\int_0^{F_1} \frac{dN}{dF}F^2 dF\propto F_1,
\end{equation}
where $F_1$ is the minimal flux of the detected source. For XMM
$F_1\sim$ 100 counts. It is convenient to relate $P_{faint}$ to 
the combined contribution of bright detected sources:
\begin{equation}
P_{bright}\propto\int_{F_1}^{F_2} \frac{dN}{dF}F^2 dF\propto F_2,
\end{equation}
provided $F_2\gg F_1$, where $F_2$ is the flux of the brightest
detected source (assuming that the same slope of LogN-LogS is
applicable to the entire range from $F_1$ to $F_2$). For XMM-Newton
$F_2 \sim$ 3000 counts. Thus the contribution of unresolved sources
can be estimated to be at the level of $\displaystyle
\frac{F_1}{F_2}\sim 0.03$ relative to the known contribution of bright
sources to the power spectrum. We used the difference between the
power spectra with and without sources to calculate the power spectrum
associated with bright sources $\displaystyle P_{bright}=P_{with
  src}-P_{no src}$ and multiplied $\displaystyle P_{bright}$ by
0.03. The subtraction of this extra component modifies the
high frequency side of the power spectrum (see Fig.\ref{fig:prsrc}),
but the magnitude of the modification does not exceed
  the uncertainties in the power spectrum, associated with pure
  Poisson noise. Given the uncertainties in estimating this component
  and taking into account that it is mostly important in the range of
  wavenumbers where Poisson noise dominates the signal, in the analysis below we neglect this component.

\subsection{Correction of the Coma power spectrum for PSF}
\label{sec:psf}
Analysis of the images containing bright point sources can also be
used to evaluate the effective power spectrum of the telescope
PSF. Indeed, the convolution of the true sky image with the telescope
PSF is equivalent to the multiplication of the true sky power spectrum
by the PSF power spectrum. Note that the PSF shape may vary across the
instrument Field of View (FoV), so one needs to perform the analysis separately for
different parts of the FoV, or use some effective PSF (e.g. FoV
averaged). We decided to follow the second approach and simply used
the observed power spectra for the Lockman Hole (Chandra) and X-Comae
region (XMM-Newton) as the PSF power spectrum (see
Fig.\ref{fig:lockps},\ref{fig:xsrc}) approximated with the following simple analytic models:
\begin{eqnarray}
P_{PSF,Chandra}=\frac{1}{\left [1+\left( \frac{k}{0.06}\right)^2\right ]^{1.1}} \\
P_{PSF,XMM}=\frac{1}{\left [1+\left( \frac{k}{0.02}\right)^2\right ]^{1.6}},
\label{eq:psf}
\end{eqnarray}
for Chandra and XMM respectively, where the wavenumber $k$ is in units of $\rm
arcsec^{-1}$. The correction for PSF blurring is done by dividing
the power spectrum of the raw image by the power spectrum of the PSF. 

\subsection{Removing the global surface brightness profile}
\label{sec:div}
\begin{figure*}
\plotwide{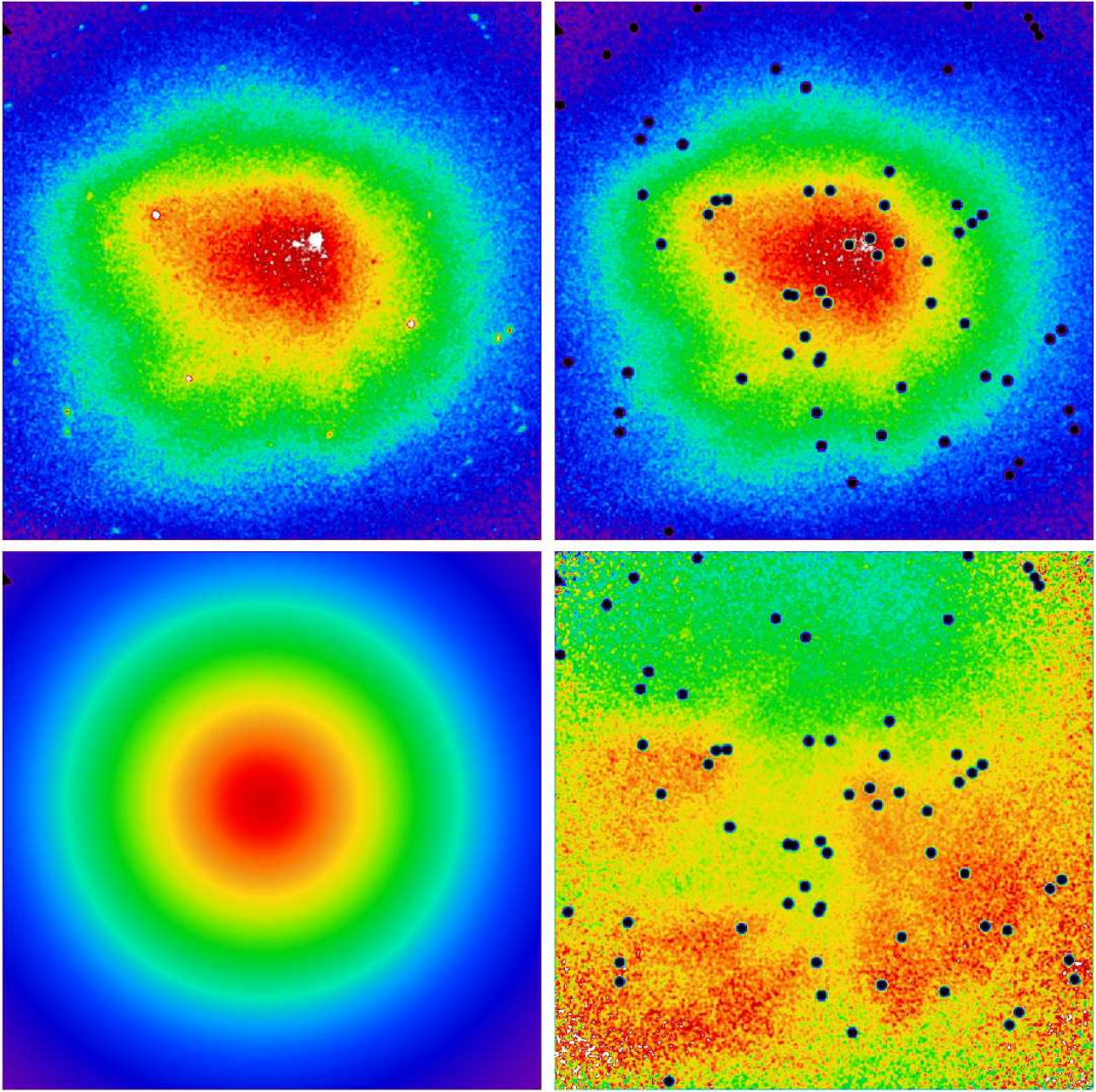}
\caption{Top-left: raw XMM-Newton image; Top-Right: image with $15''$
  circles around bright point sources removed; Bottom-Left: surface
  brightness for a $\beta$-model with $\beta=0.6$
and core radius $r_c=245$ kpc; Bottom-Right: XMM-Newton image divided by the
$\beta$-model. All images are $24' \times 24'$ (or 648$\times$648
kpc). Images are slightly smoothed with a Gaussian filter.
\label{fig:images}
}
\end{figure*}
As discussed in \S\ref{sec:3d2d} it is convenient to divide the surface
brightness of the raw image by the predictions of a simple model,
describing the global properties of Coma. The divided image can then be
interpreted as fluctuations of the surface brightness relative to
the undisturbed image. To this end we experiment with using a simple symmetric
$\beta$-model (see \S\ref{sec:data}) as an approximation of the cluster
global surface brightness distribution or a slightly more
sophisticated two-dimensional $\beta$-model, which allows for two
different core radii and rotation. For illustration the image
divided by the symmetric $\beta$-model is shown in the bottom-right corner of
Fig.\ref{fig:images}. In practice both symmetric and two-dimensional
$\beta$-models yield similar power spectra, except at the lowest
wavenumbers (see \S\ref{sec:simple}).

\begin{figure}
\plotone{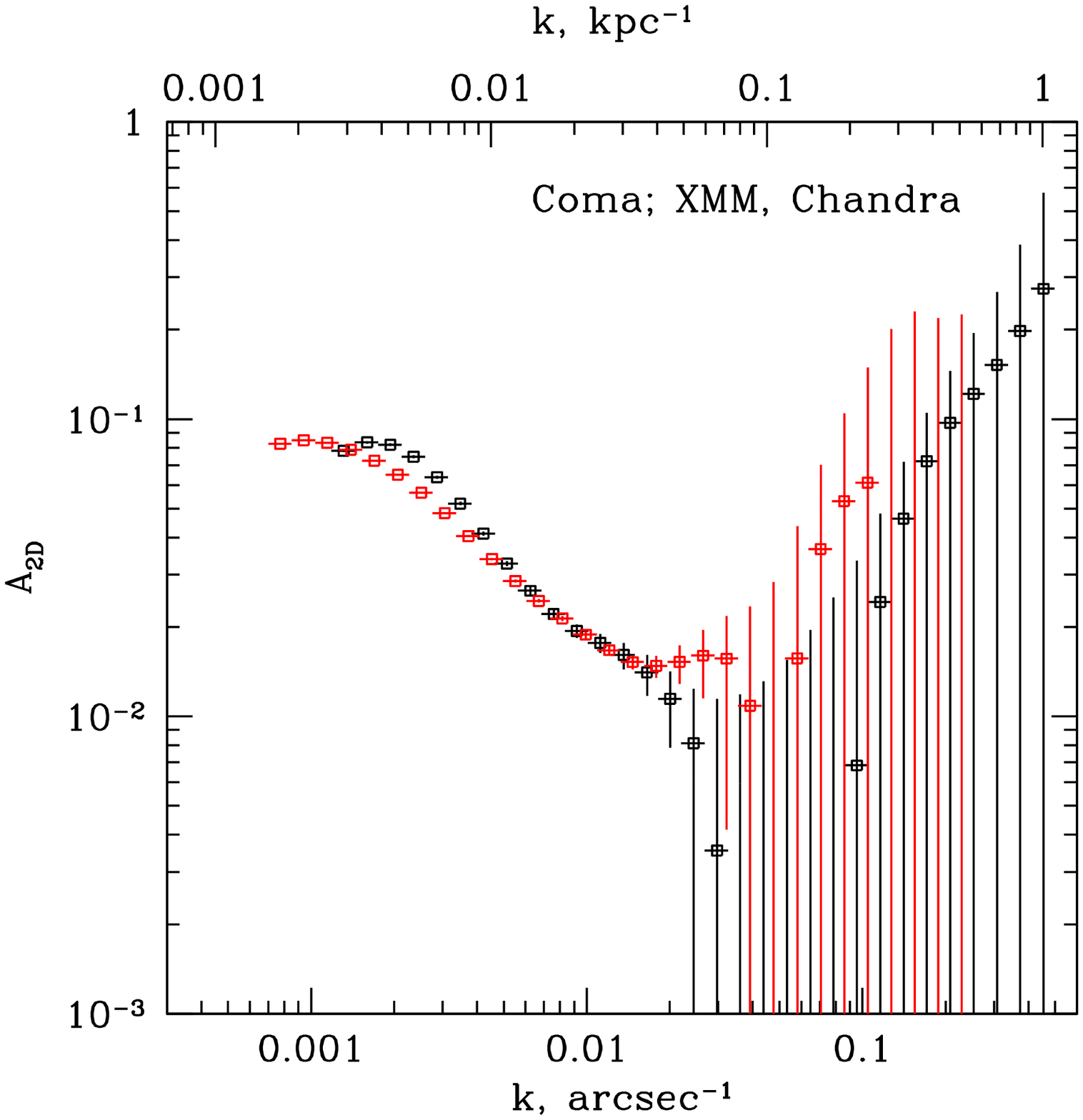}
\caption{2D amplitude of X-ray brightness fluctuations  (relative to the underlying
  $\beta$-model) as a function of the wavenumber. Black - Chandra, red -
  XMM-Newton. The uncertainties shown are due to Poisson noise only and
  do not account for the stochastic nature of density fluctuations. The
  PSF correction (see \S\ref{sec:psf}) is taken into account.
\label{fig:a2d}
}
\end{figure}

The resulting 2D power spectra (normalized by the symmetric $\beta$-model, with
point sources excised, Poisson noise level subtracted) are shown in Fig.\ref{fig:a2d}. In this plot the 2D power
spectrum was converted into the characteristic amplitude of the surface
brightness fluctuations relative to the smooth model. The measured
amplitude of perturbations of the surface brightness (before the PSF
correction) varies from $\sim$10\% at scales of $\sim$500 kpc down to
less than 1\% at scales of  $\sim$30 kpc. 

\subsection{Converting to 3D and the amplitude of density fluctuations}
\begin{figure}
\plotone{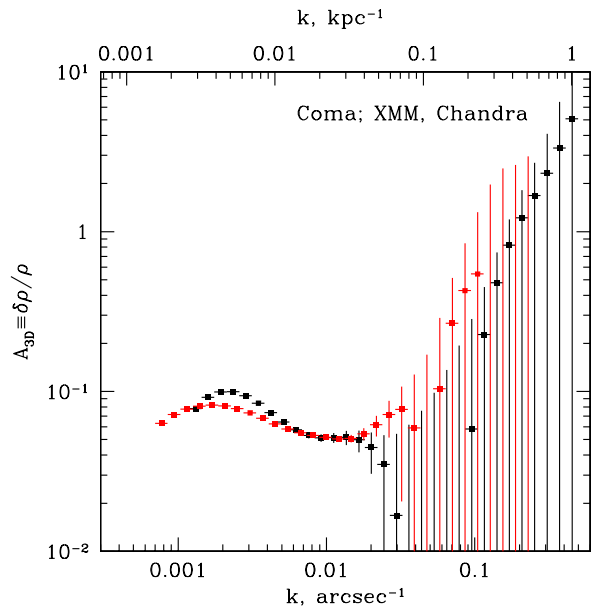}
\caption{3D amplitude of density fluctuations (relative to the underlying
  $\beta$-model) as a function of the wavenumber. Black - Chandra, red -
  XMM-Newton. The uncertainties shown are due to Poisson noise only and
  do not account for the stochastic nature of density fluctuations. The
  effect of the PSF has been taken into account in this plot.
\label{fig:a3d}
}
\end{figure}

The final step of the analysis is the conversion of the 2D surface
brightness fluctuations (relative to the underlying undisturbed image)
to the 3D density fluctuation power spectrum or amplitude of the 3D
density fluctuations. As described in \S\ref{sec:3d2d}, the
easiest way of obtaining the 3D power spectrum is to use the approximate
relation between the 3D and 2D spectra according to
eq.(\ref{eq:klarge}), which for model power spectra works well
  at wavenumbers of order $10^{-3}~{\rm kpc^{-1}}$ and larger. It is
also convenient to convert the final power spectrum to the amplitude
of the density perturbations using eq.(\ref{eq:a3}). The final
results of this analysis are
shown in Fig.\ref{fig:a3d}. This 3D amplitude characterizes typical
variations of the density relative to the unperturbed value
(i.e. $\displaystyle \frac{\delta\rho}{\rho}\equiv A_{3D}$) as a function of the spatial
 scale. The amplitude varies from $\sim$10\% at scales of a  few 100 kpc
 down to $\sim$5\% at scale of a few tens of kpc. At even smaller scales,
 pure statistical errors due to Poisson noise increase dramatically 
 precluding tight constraints on the amplitude of the
 fluctuations.

\section{Discussion}
\label{sec:discussion}

The main conclusion one can draw from Fig.\ref{fig:a3d} is that there are
significant variations of the density fluctuations of the ICM on
spatial scales ranging from $\sim$500 kpc down to $\sim$30 kpc at the
level of 5 to 10\%. These fluctuations are relatively small over
  the entire range of wavenumbers where the
existing data have sufficient statistics to make accurate
measurements. Large-scale fluctuations are directly visible in the
images divided by a suitable smooth model as excesses or dips in the
surface brightness (Fig.\ref{fig:images}), although their appearance does vary depending on
the type of the smooth model. Small-scale perturbations are in the
regime dominated by the Poisson noise and they are revealed only in a
statistical sense and not individually visible in the image.

There are several plausible physical mechanisms which can lead to the
observed emissivity/density fluctuations. Among them are the following possibilities:
\begin{itemize}
\item too simplistic model of the unperturbed gas density distribution;
\item perturbations of the gravitational potential on top of a smooth
  global potential of the cluster;
\item entropy fluctuations caused by infalling low entropy gas or by
  advection of gas from one radius to another by gas motions;
\item variations of gas density and/or pressure associated with gas motion
  and sound waves;
\item metallicity variations;
\item presence of non-thermal and spatially variable components
  (bubbles of relativistic plasma and magnetic fields).
\end{itemize}

We now briefly discuss these possibilities.

\subsection{Simplistic smooth model}
\label{sec:simple}
\begin{figure}
\plotone{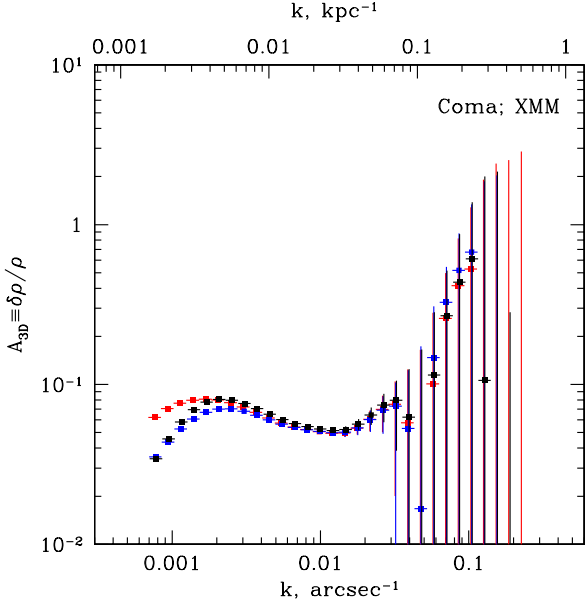}
\caption{3D amplitude of density fluctuations relative to different
  underlying $\beta$-models (based on XMM-Newton data). The red points
  (the same as shown in Fig.\ref{fig:a3d}) correspond to a spherically
  symmetric  $\beta$-model with $\beta=0.6$ and $r_c=9'$. The blue
  points correspond to a two-dimensional
$\beta$-model with $\beta=0.62$ and $r_{c,x}=9'$ and $r_{c,y}=8'$. The
  black points are for a model with $\beta=1.03$ and $r_{c,x}=15.9'$
  and $r_{c,y}=13.5'$. Depending on the parameters of the model the
  amplitude changes by a factor of 2 at scales above 500 kpc. At scales
  smaller than 300 kpc the power spectrum is virtually unchanged.
\label{fig:global}
}
\end{figure}

One of the principal uncertainties in calculating deviations of the
surface brightness or density fluctuations lies in specifying the
smooth underlying model. Our experiments with a two-dimensional
$\beta$-model have shown that varying the center of the model, the
core radii in two directions and $\beta$ within reasonable limits one can
change the amplitude of the longest perturbations (400-500 kpc) by a
factor of two, compared to a symmetric $\beta$-model (see
Fig.\ref{fig:global}). To illustrate this statement three models
  were used in Fig.\ref{fig:global}: (i) spherically
  symmetric  $\beta$-model with $\beta=0.6$ and $r_c=9'$; (ii) two-dimensional
$\beta$-model with $\beta=0.62$ and $r_{c,x}=9'$ and $r_{c,y}=8'$ and (iii)
another two-dimensional model with $\beta=1.03$ and $r_{c,x}=15.9'$
  and $r_{c,y}=13.5'$ (from \citealt{1994ApJ...435..162V}). The
  variations of the amplitude at large scales is not
surprising, since with more complex models we are able to
attribute some of the structures to the underlying model, rather than
to perturbations. The amplitude of the perturbations at scales
smaller than $\sim$300 kpc remains unchanged. In spite of the
model-dependent variations of the amplitude at large scales, the
structures seen in the images divided by the $\beta$-model are
complicated enough, so it is probably fair to state that further
reducing the amplitude of these structures would require substantially
more complex models. We concluded that a conservative systematic
uncertainty -- a factor of order of 2 -- should be attributed to large-scale perturbations, while at scales smaller than 300 kpc the
systematic uncertainty is smaller.

\subsection{Perturbations of the gravitational potential}
\label{sec:pot}
\begin{figure}
\plotone{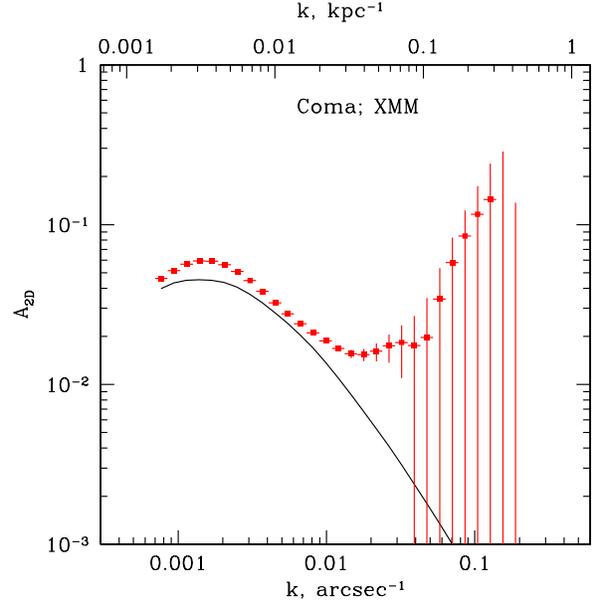}
\caption{2D relative amplitude of the surface brightness fluctuations
  (black line) caused by the presence of two components (representing
  gas concentration around two massive cD galaxies in Coma) on top of
  the global cluster emission. Parameters of the $\beta$-models for
  each galaxy and for the main cluster are taken from
  \citet{1994ApJ...435..162V}. Note that the decomposition of the
  surface brightness into the three-component $\beta$-model is not
  unique. The same set of exclusion regions (around compact sources,
  including nuclei of NGC~4874 and NGC~4889) as for real data was
  used for the simulated dataset. For comparison, the 2D amplitude
  measured from XMM-Newton data is shown with red crosses.
\label{fig:pot}
}
\end{figure}
The gravitational potential of the cluster is affected by the presence
of subhalos. In particular, in Coma there are two large cD galaxies --
NGC~4874 and NGC~4889, with the X-ray surface brightness centroid lying in
between these two objects. One can identify enhancements of the
surface brightness at the positions of NGC~4874 and NGC~4889
\citep{1994ApJ...435..162V}, although the exact amplitude of these
enhancements depends on the assumptions of the underlying cluster
model\footnote{Note that when calculating power spectra the $15''$ circles
around NGC~4874 and NGC~4889 have been excluded. Therefore the
contributions of an AGN in NGC~4889 and cool coronae
\citep{1994ApJ...435..162V} in both galaxies are excluded.}.

We can write the gravitational potential of the cluster as:
\begin{eqnarray}
\varphi=\varphi_0+\Sigma_i\Delta\varphi_i,
\end{eqnarray}
where $\varphi_0$ is the smooth potential and $\Delta\varphi_i$ is the
perturbation, associated with the subhalo number $i$.  To simplify
  the estimates we
assume that for each of the two bright galaxies their potential can be
written in an isothermal form over the interesting range of radii:
\begin{eqnarray}
\Delta\varphi=v_c^2\ln r + C,
\end{eqnarray}
where $v_c$ is the circular speed and $r$ is the galactocentric
radius, $C$ is an arbitrary additive constant. Further assuming that the gas is isothermal and ignoring
possible effects of the galaxy motion with respect to the ICM, the
relative gas
density perturbation, induced by $\Delta\varphi$ is 
took out the first minus sign in the equation but perhaps you prefer
the other way
\begin{eqnarray}
\frac{\delta\rho}{\rho}=e^{\frac{-\mu m_p \Delta\varphi}{kT}}-1=\left (
\frac{r}{r_s} \right )^{\frac{-\mu m_p v_c^2}{kT}}-1,
\end{eqnarray}
where $\mu\approx 0.61$ is the mean particle atomic weight, $m_p$ is
the proton mass, $T$ is the ICM temperature, $k$ is the Boltzmann
constant and the overall normalization is expressed in terms of
  the scaling radius, $r_s$ and the normalization constant $C$ is
  chosen so that $\Delta\varphi(r_s)=0$. For NGC~4874 and NGC~4889,
  \citet{2010MNRAS.407L..26C} measured the line-of-sight velocity
  dispersion $\sigma$ of $\sim
    283~{\rm km~s^{-1}}$ and $\sim
    266~{\rm km~s^{-1}}$ respectively at a distance of  50-60 kpc from their centers. For a spherical galaxy with
    isotropic stellar orbits a reasonable estimate of the circular
    speed is then $v_c\sim \sqrt{3}\sigma \sim 460-490 ~{\rm km~s^{-1}}$. 
 Conservatively assuming $v_c=450$ km/s and 
  the ICM temperature $kT=7$ keV,
the resulting 
radial dependence of the density perturbation is
$\displaystyle \frac{\delta\rho}{\rho}\propto \left (\frac{r}{r_s} \right )^{-0.18}-1$. Further assuming that the
perturbation of the potential extends up to $r_s=$150 kpc and is zero
at larger distances, the {\rm RMS} of the density variations induced by a single perturbation of
the potential of this type can be estimated as follows:
\begin{eqnarray}
\left (\frac{\int_0^{r_s}\left (\frac{\delta\rho}{\rho}\right )^2 4\pi r^2
  dr}{\int_0^{r_{max}} 4\pi r^2 dr}\right )^{1/2}\approx 0.03,
\end{eqnarray}
where $r_{max}\sim 325$ kpc is the characteristic size of the region
which we analyze. To simplify this estimate we have assumed that the
perturbation is located at the cluster center.
 This is a crude estimate, but
it suggests that adding a single perturbation in a form of an
isothermal halo with the above parameters is able to induce a volume
averaged RMS of order 3\%. 

Another possibility to make this estimate is to fit the observed
enhancements in the surface brightness around positions of the two
bright galaxies and use the best fitting parameters to estimate the
RMS. To this end we used a decomposition of the X-ray
surface brightness into three $\beta$-models from 
\citet{1994ApJ...435..162V} - one for the main cluster
and one for each of the two cD galaxies. Repeating the analysis for the
three $\beta$-models image divided by the $\beta$-model for the
cluster, we obtained a scale dependent relative amplitude of the surface
brightness fluctuations induced by these galaxies (see
Fig.\ref{fig:pot}). For comparison we show the density fluctuations
derived from the XMM-Newton image. Clearly, the amplitudes in the range of
scales of order few 100 kpc are in the ballpark
of values predicted by this simple exercise. This suggests that
perturbations of the potential can be responsible for a significant part
of the observed perturbations.

\subsection{Turbulence}
\label{sec:turb}
Turbulence is another possible source of pressure/density fluctuations
in the ICM \cite[e.g.][]{2004A&A...426..387S}. Assuming an eddy with characteristic velocity
$v_{eddy}$ and adiabatic behavior of the gas one can relate the
velocity  and density perturbations as
\begin{eqnarray}
\frac{\delta \rho}{\rho}\sim
\frac{v_{eddy}^2}{2}\frac{1}{\gamma}\frac{\mu m_p}{kT}\sim M^2,
\label{eq:turb}
\end{eqnarray}
where $\displaystyle \gamma=5/3$ is the gas adiabatic index, $\mu\sim 0.61$ is
the particle mean atomic weight, $m_p$ is the proton mass and $M$ is
the Mach number.
For $kT=7$ keV, the value of $v_{eddy}=450 {~\rm km~s^{-1}}$ will
induce 5.5\% variations in the ICM density.
Approximately (up to a
factor of order unity) this corresponds to the ratio of the kinetic
and thermal energies of the ICM. According to numerical simulations
this is a perfectly reasonable value \citep[e.g.][]{2003AstL...29..783S,2009ApJ...705.1129L}.  It is not clear however if
numerical simulations can accurately predict the characteristics of
the turbulent eddies (size and velocities), since the Reynolds numbers
achieved so far are not very high (and we in fact do not know the
effective viscosity of the ICM). 

The assumption that a significant part of the observed
  fluctuations is due to perturbations of the gravitational potential
  (see \S\ref{sec:pot}) would imply that the level of turbulence is
  modest. For instance one can use Fig.\ref{fig:pot} to estimate the
  amplitude of fluctuations on top of the contribution induced by gas
  concentration around the two massive cD galaxies. These excess
  fluctuations of density have an amplitude of order 2-4\% on scales
  $\sim$30-300 kpc, which corresponds to a velocity amplitude of
  $\sim 300-400 {~\rm km~s^{-1}}$ according to eq.\ref{eq:turb}. This
  in turn implies that the contribution of turbulent pressure support is
  similarly low. Accounting for the gas entropy fluctuations
  considered in section \S\ref{sec:entropy} may lower this estimate
  even further. Of course accurate estimates of the turbulent pressure
  support by this method are difficult.  Future observations with
X-ray micro-calorimeters, initially from ASTRO-H
\citep[e.g.][]{2010SPIE.7732E..27T}, should be able to constrain both
the velocities and sizes of the energy-containing eddies
\citep[e.g.][]{iz11}. For now, turbulence remains a plausible
contributor to the observed surface brightness fluctuations, as well
as pressure fluctuations \citep{2004A&A...426..387S} in Coma.

\subsection{Sound waves generated by the turbulence}
It is interesting to estimate a possible contribution of sound waves
to the observed surface brightness variations. The advantage of sound
waves is that they represent a  
linear relation (rather than quadratic) between the gas
velocity and the pressure/density perturbation:
\begin{eqnarray}
\frac{\delta \rho}{\rho}\sim \frac{v}{c_s},
\end{eqnarray}
where $c_s$ is the sound speed. E.g. to produce 5\% variations in
density one needs $v\sim 70 {~\rm km~s^{-1}}$. Sound waves cross the
region of interest $R\sim 350$ kpc over $t_s=R/c_s$. If one assumes
that the sound waves are generated by the turbulence, then the rate of
sound-wave generation per unit volume (${\rm ergs~s^{-1}~cm^{-3}}$)
is \citep[][ \S 75]{1952RSPSA.211..564L,1987hydr.book.....L}
\begin{eqnarray}
\dot{\epsilon}\propto \frac{v_{eddy}^8}{c_s^5 l},
\end{eqnarray}
where $l$ is the characteristic eddy size. The typical amplitude of
the gas velocity due to sound waves in the region of size $R$
(assuming that turbulence generates sound waves uniformly over the
region with volume $\sim R^3$) can be estimated as 
\begin{eqnarray}
v\approx \sqrt{\dot{\epsilon}t_s}=v_{eddy}\left ( \frac{v_{eddy}}{c_s} \right
)^3\sqrt{\frac{R}{l}}\sim 30 {~\rm km~s^{-1}},
\end{eqnarray}
where we assumed $v_{eddy}=450 {~\rm km~s^{-1}}$ and $l=100$ kpc. Thus sound
waves generated by turbulence, are not very effective in
producing density perturbations unless the turbulent velocities
generating the sound waves are larger than $450 {~\rm km~s^{-1}}$. Of
course the above argument provides only an order of magnitude estimate. 

\subsection{Entropy variations}
\label{sec:entropy}
Another natural mechanism, which may cause density variations is the
variation of entropy/temperature. We distinguish them from potential
perturbations, turbulent pulsations or sound waves, since entropy
variations can exist without associated strong pressure perturbations
(to lowest order in Mach number). We assume that in these perturbations
\begin{eqnarray}
\frac{\delta \rho}{\rho}\sim -\frac{\delta T}{T},
\end{eqnarray}
where $T$ is the gas temperature.

One can identify two different modes of entropy perturbations. The
first mode is associated with the low entropy gas, infalling into the
system. One of the features, seen in the Coma images, namely a chain
of galaxies apparently infalling into Coma from the East
\citep{1997ApJ...474L...7V} is associated with colder gas, which is
clearly seen in the hardness ratio map \citep[e.g.][]{2003A&A...400..811N}. In
Fig.\ref{fig:images} this cold stream appears as an enhancement
in the surface brightness relative to the cluster $\beta$-model. The
amplitude and the area covered by this feature are comparable with the
perturbations, identified as a potential perturbation in
\S\ref{sec:pot}. We therefore can expect a similar impact of this
feature on the final estimate of the density perturbations.

Another possibility to create entropy fluctuations is the advection of
gas with different entropies by gas motions in turbulent
eddies. Assuming an unperturbed isothermal atmosphere of the cluster
with density following a $\beta$-model, we can estimate the eddy
size needed to create the required density contrast. When a gas lump
is advected from the radius $r$, where the gas pressure is $P$, to the
radius $r+\delta r$, where the pressure is $P-\delta P$, and it
expands adiabatically, the corresponding density change is
\begin{eqnarray}
\frac{\delta \rho}{\rho}=\frac{1}{\gamma}\frac{\delta P}{P}\approx -\frac{1}{\gamma} 3\beta \left ( \frac{\delta r}{r} \right )\frac{\left ( \frac{r}{r_c}\right )^2}{1+\left ( \frac{r}{r_c}\right )^2}.
\end{eqnarray}
Near the core radius $r\approx r_c$ a 10\% change in radius causes a
$\sim$5\% change of density. Thus eddies of size $\delta r\sim 0.1 r_c\sim
25-30~$kpc are needed. 
As in the case of turbulent fluctuations, future measurements of
eddy amplitudes and sizes would provide crucial constraints on the
contribution of this mode.

\subsection{Metallicity variations}
\label{sec:metalls}
In the 0.6-2.5 keV band the change of metallicity from 0.5 solar to 1
solar (for $kT=8$ keV APEC  model, \citet{Smi01}) causes a 5.6\% increase in the count rate
(for XMM-Newton). The change of density, causing the same increase of
the flux is 5.6/2=2.8\%. Thus very strong variations of metallicity
are needed if they are to make a significant contribution to variations of
emissivity.

\subsection{Bubbles of relativistic plasma}
\label{sec:bubbles}
Observations of cool-core clusters revealed rich substructure caused
by the presence of AGN-inflated bubbles of relativistic plasma
\citep[e.g.][]{2000A&A...356..788C}. In terms of X-ray surface 
brightness perturbations, 
the bubbles reveal themselves as depressions in the surface
brightness with the amplitude set by the fraction of 
the line-of-sight volume occupied by the bubbles. In cool-core clusters,
the fraction of the volume occupied by bubbles can be large - of order
10\% or more. The RMS variations of the density perturbations
will be high accordingly. In Coma there is no compelling evidence for
widespread bubbles of relativistic plasma, although one of the cD galaxies -
NGC4874 does contain an active radio source with a pair of bubbles on
each side of the nucleus \citep{2005ApJ...619..169S}.
Future low frequency
radio observations might reveal radio bright regions (associated with
low energy electrons contained in bubbles). If these regions correlate
with the X-ray surface brightness depressions, this would be a key
indicator in favor of this mechanism. For now we can only state that
in Coma, radio bubbles are a possible source of X-ray surface
brightness fluctuations, although there exists no compelling evidence that
this mechanism is really important for Coma.

\section{Conclusions}
X-ray surface brightness fluctuations in the core ($650 \times 650$
kpc) region of the Coma cluster are analyzed using XMM-Newton and
Chandra images and converted into scale-dependent amplitudes of volume-filling ICM density perturbations. The characteristic amplitudes
relative to a smooth underlying density are modest: they vary from 7-10\%
on scales of 500 kpc down to $\sim$5\% at scales $\sim$ 30
kpc\footnote{for 
comparison the mean free path for Coulomb collisions in
the core of Coma is of order 5 kpc}. At smaller
scales,  projection effects and the PSF (for
XMM-Newton) smear the density fluctuations by a large factor,
precluding strong limits on the density fluctuations in
3D. Nevertheless making deep ($\sim$ Msec long) Chandra exposures can
still be useful in pushing down the constraints for scales smaller
than 30 kpc.

Several physical effects can contribute to observed fluctuations of
the surface brightness/density. The most likely contributors at large
scales are the perturbations of the gravitational potential by massive
cD galaxies and entropy variations due to infalling cold gas. We do see in
raw images the substructure which can be plausibly associated with
these mechanisms. 

Other effects can contribute to the observed fluctuations, including
variations of density in turbulent eddies, advection of the low/high
entropy gas by turbulent eddies, generation of sound waves by 
turbulence, bubbles of relativistic plasma and metallicity
variations. Future high energy resolution X-ray missions
(e.g. ASTRO-H) and low frequency radio observations (e.g. LOFAR) could
help differentiate between these possibilities. At present we can
consider the detected variations as conservative limits on the
amplitude of the density/emissivity fluctuations caused by any of
these effects.

Despite a variety of possible origins for density fluctuations, the
gas in the Coma cluster core is remarkably homogeneous on scales from
$\sim$ 500 to $\sim$30 kpc, with density variations relative to
a plausible smooth model smaller than 10\%.

\label{sec:conclusions}

\section{Acknowledgments} 
This research has made use of data obtained from the Chandra and
XMM-Newton Archives.
This work was supported in part by the Leverhulme Trust Network for
Magnetized Plasma Turbulence and STFC Grant ST/F002505/2 (AAS).The
work was supported in part by the Division of Physical Sciences of the
RAS (the program ``Extended objects in the Universe'', OFN-16). The
financial support for SWR was partially provided for by the Chandra
X-ray Center through NASA contract NAS8- 03060, and the Smithsonian
Institution.

\label{lastpage}
\end{document}